# Complexity Synchronization in Emergent Intelligence

Korosh Mahmoodi[1*], Scott E. Kerick[1], Piotr J. Franaszczuk[1,2],
Thomas D. Parsons[3], Paolo Grigolini[4], Bruce J. West[4,5]

[1*]US Army Combat Capabilities Development Command, Army Research Laboratory, Aberdeen Proving Ground, 21005, MD, USA.
[2]Department of Neurology, School of Medicine, Johns Hopkins University, Baltimore, 21287, MD, USA.
[3]Computational Neuropsychology Simulation Laboratory, Edson College, Arizona State University, Tempe, 85004, AZ, USA.
[4]Center for Nonlinear Science, University of North Texas, Denton, 76203, TX, USA.
[5]Office of Research and Innovation, North Carolina State University, Raleigh, 27695, NC, USA.

*Corresponding author(s). E-mail(s): koroshmahmoodi@gmail.com;
Contributing authors: scott.e.kerick.civ@army.mil;
piotr.j.franaszczuk.civ@army.mil; thomas.parsons@asu.edu;
paolo.grigolini@unt.edu; brucejwest213@gmail.com;

**Abstract**

In this work, we use a simple multi-agent-based model (MABM), implementing selfish algorithm (SA) agents, to create an adaptive environment and show, using modified diffusion entropy analysis (MDEA), that the mutual-adaptive interaction between the parts of such a network manifests complexity synchronization (CS). CS has been experimentally shown to exist among organ-networks (ONs) of the brain (neurophysiology), lungs (respiration), and heart (cardiovascular reactivity) and to be explained theoretically as a synchronization of the multifractal scaling parameters characterizing each time series. Herein, we find the same kind of CS in the emergent intelligence (i.e., without macroscopic control and based on self-interest) between two groups of agents playing an anti-coordination game, thereby suggesting the potential for the same CS in real-world social phenomena and in human-machine interactions.

**Keywords:** Complexity Synchronization, Adaptive Environment, Emergent Intelligence, Multi-Agent-Based Modeling, Selfish Algorithm, Reinforcement Learning, Modified Diffusion Entropy Analysis

## 1 Introduction

We explore the potential connection between complexity synchronization (CS) phenomena recently observed in the interaction among the time series generated by human heart, lungs and brain [1, 2] with that of phenomena generated by the interaction of correspondingly relative complexity of social Selfish Algorithm agents (SA-agents) of [3]. Note that the detailed dynamics of the two kinds of networks, the first one being the interaction of three human organ-networks (ONs), whereas the second is the interaction of an arbitrary number of SA-agents, can be very different, and yet we shall show that the multifractal nature of the scaling indices of their time series satisfies the conditions for CS. The equivalence of the information content of the human physiological and social networks is established dynamically by the direct calculation of the collective behavior of the SA-agents.

This CS of the scaling indices of the SA-agent interacting groups forces the conclusion that above the level of time series synchronization [4, 5] there is a higher-order synchronization of the time-dependent scaling indices [2]. This higher-order synchronization can arise even in the absence of synchronization of the central moments of the underlying time series, such as in their cross-correlation function. But to see this behavior clearly requires laying the foundation for the growth and formation of a network from the mutual interactions of a collection of SA-agents, bearing in mind that the dynamics of such self-serving agents leads to the counter-intuitive result that the social network that is so formed derives optimum value from their interactions [3].

Each SA-agent learns to modify its model of the world to anticipate/engineer its dynamic environment (i.e., the behavior of other SA-agents), whose purpose is to increase its own payoff. In this model, to make decisions (i.e., in which direction to move), the SA-agent first adaptively decides to either rely on the information from neighboring SA-agents of the in-group or out-group neighboring SA-agents (using a reliance propensity/probability). In the first case, the SA-agent picks its next direction to move as the average of those in-group agents. Otherwise, using the information from the neighboring out-group members of the other SA-agent group, the SA-agent anticipates their average position and adaptively decides what its next direction to move will be, either away from or towards that position (using a oppose-follow propensity/probability).

Starting from random positions and moving in random directions on a two-dimensional lattice with periodic boundary conditions, thereby walking on the two-dimensional surface of a three-dimensional torus in phase space, the SA-agents, each trying to increase their own payoff, learn to form multiple large collectives, such as swarms. Although the bulk of the discussion presented herein is in terms of the numerical simulation of social interactions among selfish individuals focusing on the resulting emergence of group intelligence we do not want to leave the impression that our remarks are constrained to such applications. For example, if wealth is a measure of social value, so that in reaching their goal of individual wealth maximization, they incidentally and counter-intuitively also optimize the wealth of the social group. An analogous argument applies to collections of other animals as well as to physiologic network of organ-networks (NoONs).

The intelligent behavior of animal collectives is one research focus of Couzin [6], who, using an integrated approach involving both fieldwork and laboratory experiments guided by mathematical models and computer simulations, attained unparalleled understanding of animal group behavior. For example, in his remarkable work with army ants he unveiled how such ants form traffic lanes in their movement that is optimum for the avoidance of congestion. The social value in this latter case being the efficiency of movement to carry out a group task.

The intelligence of our multi-agent-based model (MABM) society is emergent since there are no macroscopic control parameters, such as noise or temperature. The first is often used in swarm intelligence models [7], and the latter (temperature) is often used in the social application of the Ising model of phase transitions in which the social model has the control 'temperature' being interpreted as the level of excitation of the social network [8]. In the present MABM society, the adaptive environment of each individual SA-agent replaces the role of the macroscopic control parameter. Proceeding in this vein, the ensemble average of reliance propensity, oppose-follow propensity, and the order parameter time series of two SA-agent-groups represent the dynamics of the mutual adaptive interactions in the social network. If $X(t)$ is the time series of interest and it scales in such a way that given the constant $\lambda$ we obtain the homogeneous scaling function $X(\lambda t) = \lambda^\delta X(t)$ enabling us to measure the time series complexity by means of the modified diffusion entropy analysis (MDEA.)

The MDEA reveals that the scaling indices $\delta$'s of these time series are time-dependent and in synchrony with one another, meaning their statistics change over time relative to their changing environment. This means that each scaling index is related to a multifractal dimension, and the CS is expressed in terms of the synchrony of these multifractal dimensions. We demonstrate that although the cross-correlation between these time series does not reveal the synchronous interrelations among the groups, there is strong cross-correlation between their multifractal time series. This cross-correlation is a CS and provides a reasonable measure of the intelligence of the overall network.

## 1.1 Temporal Complexity

The systemic environment a complex network is comprised of other complex networks, which creates a network-of-networks (NoNs), that is either competing or cooperating. As such, for a NoNs to carry out a specific task, each interacting network should be mutually adaptive to its environmental changes. For example, the environment of a particular organ-network (ON) consist of the other adaptive organ-networks within a NoONs. It has been shown that the time series generated by the internal dynamics of these separate ONs, e.g., two such ONs in the human body generate the respiration and cardiac datasets, both of which have temporal complexity [9, 10], meaning that their statistics are governed by Crucial Events (CEs).

CEs are defined by discrete time series consisting of events with time intervals $\tau$ between consecutive events that are statistically independent of one another. The defining properties of renewal events (RE) were put together by William Feller, a giant in the field of stochastic processes, who introduced the name Renewal Theory in his 1950 two-volume *magnum opus* [11], wherein he developed the mathematical foundation of RE. The first property of CEs is therefore that they are, like Poisson processes, REs.

However, although CEs are renewal they, unlike Poisson events, have an inverse power law (IPL) probability density functions (PDFs), i.e., the IPL PDF of the time intervals between sequential events, $\tau$'s, is $\psi(\tau) \propto \tau^{-\mu}$ and the IPL index $\mu$ is in the domain $1 < \mu < 3$, or equivalently, such CE time series have inverse power law spectra of the form $S(f) \propto f^{-\beta}$ where $0 < \beta = 3 - \mu < 2$ [12].

The utility of REs have been found in modeling the firing of neurons [13], to analyze the reliability and maintenance of networks modeled generally with time-dependent parameters, these and many more are considered in the brilliant work of Cox [14] and which were applied to medical phenomena in [15]. For most physiological data $2 < \mu < 3$, with $\mu \simeq 2$ in a healthy brain [9]. To unambiguously measure the IPL index $\mu$ of a single complex trajectory, the data processing technique called MDEA was developed [16], mainly because in the ergodic region of $2 < \mu < 3$ the second moment does not exist, and in the non-ergodic region of $1 < \mu < 2$ both the first and second moment do not exist, therefore, complexity analysis based on the second moment, such as detrended fluctuation analysis (DFA), can give misleading results. The MDEA applied to a fractal time series provides a scaling index $\delta \in (0, 1]$, with $\delta = 0.5$ corresponding to Brownian motion. The scaling index $\delta$ is connected to the temporal complexity IPL index $\mu$ of an ergodic signal by $\mu = 1 + 1/\delta$. In the present work, we show that the scaling index $\delta$ found in mutual adaptive networks is time-dependent, and those in NoNs are in synchrony with one another. For more details on MDEA, see [1].

## 1.2 Mutual Adaptive Environment Breeds Intelligence

The interactions between atoms, which are the building blocks of matter, are identical and non-adaptive. Contrariwise, the interactions between entities, such as individuals within a social network or a network of neurons within a brain, are non-identical and adaptive, resulting in emergent properties. So, to model such complex networks, we need building blocks that can update their interactions based on their internal state. Agents have been used for this purpose because they can receive information about their changing environment, process that information, make decisions based on the application of weighted valuations, and implement those decisions to manipulate their environment (such as sending true or deceptive information, moving, etc.).

To the best of our knowledge, the first attempt at using mutual-adaptive agents in an adaptive environment was in [17, 18], wherein each agent $i$, located on a 2D lattice, was able to choose the microscopic control parameter value ($K_i$) reflecting the extent to which it would imitate the decision of its neighboring agents. The agents could either cooperate (*C*) or defect (*D*). Based on the agent's decision and the decision of its neighbors, payoffs were received according to the payoff matrix of the prisoner's dilemma (PD) game. Each agent used an Ising-like rate function containing the parameter $K_i$ to make its decision. At the end of each trial, the agent compared its current payoff with its previous payoff and used that difference as feedback to reinforce its $K_i$ choice for the subsequent decision-making trial.

The idea of using the successive differences in payoffs as feedback and thereby bringing the network to criticality surfaced in [19, 20] wherein it is shown that the act of increasing the control parameter of the Ising-like rate towards its critical value has the remarkable effect of increasing the average payoff of the network. So, given the continuous weighting updates of the microscopic imitation parameter $K_i$ of the agents, based on their payoff feedback, the entire network reaches criticality, this phenomenon has been identified as self-organized temporal criticality (SOTC) in [17, 18]. SOTC results in a robust and resilient team of cooperators. It is important to note that SOTC is distinguished from its precursor self-organized criticality (SOC) in that SOTC does not include macroscopic control parameters (such as temperature as in the Ising model [21] or noise as in [22]). Instead, the mutual interaction between the agent and its adaptive environment modifies its microscopic control parameter $K_i$, and the network, as a whole, reaches a global understanding of the advantage of cooperation and long-range correlation. In other words, each agent, although acting in its own self-interest, contributes to the emerging network's intelligence, and such intelligence controls the behavior of agents and the system as a whole.

Criticality as the antecedent to SOC has been hypothesized to be able to describe the dynamics of biological systems such as the brain [23]. Despite some success in using criticality to explain some aspects of intelligence, it has limitations:

- The theory of criticality is typically controlled via a macroscopic control parameter rather than being self-controlled [23, 24].
- The network's units are located on a predefined lattice rather than being on one that is initially free and is dynamically formed, consequently emerging through the interactions of the units [25].
- The network at criticality is highly responsive to weak perturbations and is consequently not robust [26].
- The response of a network at criticality to a stimulus is prolonged. The process is known as *critical slowing down* [27], meaning that its dissipation is weak.

While SOTC overcomes these limitations [17, 18], it still needs a predefined two-dimensional lattice network, even though the connectedness of the network on this lattice changes over time. Also, it uses the Ising-like rules for updating the decisions. Note that an agent in SOTC does not make its decision (*C* or *D*) independently of the other agents. Instead, it decides whether to increase or decrease the biased weighting of its microscopic imitation ($K_i$) based on the decisions of its neighbors during the playing of PD.

Chialvo et al. [28, 29] suggested a long-range correlation, parameterized to the autocorrelation function, as a fundamental feature of intelligence and criticality. Using autocorrelation as feedback to update the macroscopic control parameter, they showed that the network can stay within a narrow region of criticality. Their work might be considered a top-down version of SOTC [17, 18]. However, it implies the assumption that the NoNs somehow knows that long-range correlation (LRC) is beneficial, while in SOTC, the LRC is a byproduct of bottom-up self-organization rather than being the generator of LRC. Also, it is not always true that LRC correspond to self-organization. It has been shown [30] that if the LRC is the result of long-term memory (such as in Fractional Brownian Noise), it is a sign of the organization's impending collapse rather than indicating a robust, self-organizing network at a high level of performance, as it would if generated by SOTC.

### 1.3 Payoffs as Environmental Feedback

In MABM, the payoffs of the game that SA-agents play give them feedback for their actions toward others, letting them update their weighted biases for future decisions toward optimized performance. We can consider the game as a metaphor for the fixed part of the environment in which SA-agents are co-located, such as obstacles they must overcome to succeed. On the other hand, the SA-agents' behavior changes over time, creating an adaptive environment for one another as in the NoNs.

For example, to model an environment where only cooperation results in optimal, long-term success, it is common to set the SA-agents to play the PD game. The payoffs of the PD game are as follows: When the pair of SA-agents make the decision cooperation C, each gets 1. If one SA-agent decides C and the other decides $D$, they receive 0 and $1 + T$, ($T > 0$ is the temptation to cheat), respectively. If both agents decide $D$, each receives 0. Note that $1 < 1 + T$, so agents are tempted to choose $D$ over $C$, and $1 + 1 > 1 + T + 0$, which means cooperation is a better option in the long run.

SA-agents are incentivized to change their decisions based on the feedback of the PD game's payoffs, and if they learn over time, they might find that mutual cooperation is the most beneficial decision. These payoffs can be representative of an environment where two SA-agents need to reach a super-ordinate goal, but they can only do so if one SA-agent elects to assist the other and shares the benefit (mutual $C$). But the SA-agent in the super-ordinate position may also decide to hold on to the benefit ($D$) or never choose to assist the other (mutual $D$). Over time, an SA-agent can change decisions (and so indirectly change the decisions of the other SA-agent) to reach the "eureka" moment of mutual cooperation.

### 1.4 Selfish Algorithm (SA)

To go beyond the limits of the SOTC agents in [3, 24] the selfish algorithm(SA) agent (SA-agent) is introduced to create a novel architecture for modeling emergent intelligence (EI), see [31, 32] for discussions of the variety of well-defined EIs. Here we interpret EI to be how the pattern of decisions made by the collective of interacting SA-agents produces a greater reward for the social group than would be obtained for the same group but with each member making an independent decision. The EI observed herein arises from the SA-agents making their decisions based solely on reinforcement learning; an SA-agent assigns a $P$ value for each choice and reinforces it according to the difference between its corresponding last two payoffs. In other words, an SA-agent biases its $P$s, using the feedback of its two last payoffs to anticipate/engineer the next behavior of the other SA-agents in its environment. That this will result in the suggested form of EI is not obvious and only reveals itself through the proper experimental computations.

SA-agents can make various decisions, such as *C* or *D* in the PD game, trust or not trust the decision of other SA-agents, interact or not interact with specific SA-agents, etc. The set of the $P$s of the SA-agents' choices across events creates temporally complex networks (such as a network of trust and a network of connections) that emerge among the SA-agents because of their mutually adaptive interactions. These networks accumulate what SA-agents learn through the experience of playing with one another. Note that the intelligence emerges among the SA-agents without assuming a pre-existing network structure, and instead, SA-agents form multi-layer temporally complex networks from a lattice-free initial state [24].

In [3], SA-agents play the PD game with one another and have choices C or D. In playing the PD game, SA-agents reach a level of mutual cooperation over time, depending on the value of the temptation to cheat ($T$). Adding the choice of trusting or not trusting the decision of other SA-agents, and the choice of with whom to play, SA-agents show higher levels of mutual cooperation and robustness. Emergent networks (ENs) resulting from SA-agent interactions are resilient to perturbations and maintain high performance. In [3], some SA-agents were replaced with zealots (i.e., SA-agents that only chose D) to show that the ENs maintain their functioning by isolating the zealots.

It is important to note that from time to time some SA-agents defect (to get a free ride) and receive the benefit of being in a pool of cooperative agents, but quickly, the other SA-agents react to such behavior. For example, these SA-agents will either not play with such SA-agents or will play as defectors with them. In this way, other SA-agents force the defecting SA-agents to return to cooperative behavior ([3], Figure 2.) and thereby teach the deviant SA-agent how to be a 'good citizen'. Such bottom-up control indicates a healthy, robust, adaptive social group.

The main advantage these SA-agents possess is that the emergent temporally complex networks are interpretable since they are based on fuzzy logic ($P$s) and originate from self-interest. For example, the temporally complex network of connections surrounding the zealot demonstrates that SA-agents did, in fact, isolate the zealot ( see Figure 11 in [3]). In contrast, the networks formed from traditional artificial intelligence (AI) networks, such as deep neural networks, are far less interpretable as part of the theory and have been described as "black boxes". Although the SA-agents in [3] have three types of choices, it is easy to equalize their influence on the network dynamics by adding more choices, such as deception, for real-world applications.

## 1.5 Complexity Synchronization (CS) differs from complexity matching (CM)

In [1, 2], it has been shown that the scaling of simultaneously recorded time series from physiological ONs, e.g., the brain, heart, and lungs, obtained by MDEA processing produces distinct time-dependent scaling indies. Also, such time-dependent behavior of the scaling indices of these interacting ONs are in synchrony with one another, a phenomenon we designated as complexity synchronization (CS). In the present work, we study the occurrence of the same phenomenon in simulated datasets generated from interacting SA-agents to show that CS is a general emergent property of mutually adaptive environments, whether using physiological datasets or model-generated surrogate data. Since the CS between SA-agents has the same mathematical infrastructure as that of physiological ONs, it suggests the hypothesis that SA-agents could be used as trainers, rehabilitators, decision-makers, etc., to form successful human-EI hybrid teams.

Complexity matching (CM) [33] identifies the transfer of information as being maximal when the complexity level identified by the multifractal dimensions of two interacting complex networks are the same [34]. It is a phenomenon that has been fully explained only by means of the Shannon-Wigner information entropy, i.e., the complexity of the information-rich network does not change when interacting with a information-deficient complex network, however the information-poor network slowly increases in (information) complexity until it is on a par with the information-rich network. The exchange of information does not follow that of energy and apparently differs from the traditional form of the Second Law of Thermodynamics.

When two physical networks of approximately the same size are brought into contact with one another, the hotter network loses energy to the cooler network until they reach thermal equilibrium and the temperature is uniform across the two. Information transfer does not work this way. It is true that the information-rich network increases the complexity level of the information-poor network through the transfer of information, doing so without necessarily losing any of its own complexity and therefore its information content remains the same. This kind of implementation of CM has been used to interpret the arm-in-arm walking of elders who have impaired gate patterns with younger persons to reinstate a better gait pattern [35]. Moreover, it suggests that complexity is a qualitative property of an ON, however, its measure, the multifractality, is quantitative.

By way of contrast, in CS when two or more networks strongly interact, and their complexities change synchronously. So, CM cannot by itself explain the CS phenomenon. In [36], two SA-agents were shown to strongly interact to model the experimental results of rehabilitation in patients with gait disorder when they walked arm-in-arm with a care-giver [37]. This model can lead to a theoretical foundation for CS, showing that the network with lower complexity can improve its complexity through information obtained from the more complex network.

## 1.6 CS measures EI

Here, we point out that a complex network generating a CE time series is not necessarily intelligent. For example, blinking quantum dots, liquid crystals, earthquake waves, solar flares, etc., clearly, are not intelligent, while the time series associated with their dynamics have a complexity index $\mu$ which is close to that of biological networks such as the brain [38, 39]. These processes host CEs time series that could be termed mechanical to distinguish them from the CEs time series generated by ONs, but which share the same statistical properties. On the other hand, intelligence is an emergent property, i.e., individuals contribute to realizing an achievement they could not achieve alone, and in return, the formed group's intelligence steers an individuals' behavior [3, 17]. So, an EI network continuously adapts to its changing environment to better anticipate/engineer the next state of the environment. By attaining that next state the organization obtains the advantage of being a step ahead of the decision-making in time and, consequently, improves its self-interest/survival. This means that in an EI network CEs emerge as a result of bi-directional interactions between its elements, and in addition the EI network is able to change its CEs statistics (scaling index) to that of its adaptive environment, a phenomenon that is quantified using CS. So, an EI hosts intelligent CEs.

In this work, we use MABM to create EI, and by studying the dynamics of the emerging networks, we find the CS phenomenon to be in evidence, which was only recently uncovered in the analyses of simultaneously recorded brain, heart, and lung time series datasets [1, 2]. This finding suggests a universality of CS beyond the confines of physiology and sociology and to be a property resulting from the intelligent CEs that makeup the respective time series.

## 2 Model/Methods

To generate CS in a social context, we model the competition between two groups of SA-agents. Then, using MDEA, we study the high-order cross-correlation between the multifractal time series of the scaling indices characterizing different mean fields emerging from the distinct dynamics of the two groups. Put more simply, the two time series, one for each group of SA-agents are processed to determine the behavior of their separate and distinct scaling indices $\delta$ in time which we have shown elsewhere [2, 36] also determines the multifractal dimensions of the separate time series.

### 2.1 Selfish Algorithm Agent (SA-Agent)

SA-agents are the building blocks of the EI studied herein. An SA-agent makes decisions based on the $P$s it assigns for different choices and updates them over time throughout the interaction with other SA-agents. For each SA-agent, other SA-agents act as mutual adaptations within environments, as defined previously using NoONs in a physiologic network [1, 2]. An individual SA-agent:

1. has sensors, with inputs receiving signals in the form of sound, vision, taste, tactile, or smell from its environment. The source of information can be from other SA-agents or its payoff, which SA-agent uses in the decision-making process or as feedback for reinforcing the corresponding $P$s, respectively.
2. has a way of making decisions based on what it learned from past experiences (its model of the world) and received information. Decision-making is SA-agent's way of changing its environment, such as moving (legs and hands) or sharing information (communication).
3. has a way of updating its set of $P$s (its model of the world) based on the feedback it receives.
4. has an incentive to improve its payoff/performance.

To make a decision at a given trial, the SA-agent generates a random number from a uniform distribution [0 1] that would fall in one of the intervals, which is to say, collapses the $P$s to a decision. For example, if the blue cross illustrated in Figure 1 shows the value of the random number generated by the SA-agent, then the decision of the SA-agent in this trial is 'B.' The random number that the SA-agent uses in each decision-making trial prevents the dynamics from being deterministic/fragile and helps the SA-agent sustain exploration. For example, if the value of *P* for a choice is small but non-zero, the SA-agent may still select its corresponding choice.

The decision-making process of a SA-agent (2 above) relies on adaptive *P*s. In the case of binary choice, *P* and $1 - P$ represent the propensity of the two complementary choices. In Figure 1, we demonstrated this using a moving threshold which divides the interval [0 1] into two sections, each corresponding to a choice (*A* or *B*).

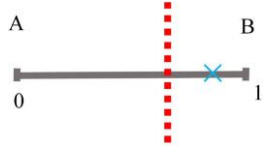

**Fig. 1** Schematics of the decision mechanism of SA-Agent for binary decision making. "*A*" and "*B*" represent the choices, and the red dotted line shows the position of the moving threshold, splitting the interval into *P* and $1 - P$. The blue cross shows the random number generated at a given trial.

Note that if there were *d* choices for a SA-agent to pick from, then $d - 1$ thresholds are needed, and the intervals between those thresholds correspond to the *P*s of the choices. Also, the SA-agent might have different choices which should be taken consecutively. For example, a SA-agent in [3] must first choose another SA-agent with whom to play the PD game, then making a choice between "*C*" and "*D*", and finally choose to trust or not to trust the decision of the other SA-agent. After decision-making, the SA-agent receives a payoff and uses that payoff as feedback to update the position of the corresponding thresholds (values of the *P*s) that contributed to the final choice (reinforcing the *P*s). The process of updating the *P*s (3 above) requires the SA-agent to:

1. save the position of the threshold(s) and the payoff of the current and previous trial.
2. use Equation 2 to reinforce (increase the *P* of the successful decision(s)).

Note that a SA-agent's model of the world, represented in complex networks of *P*s, is based on fuzzy logic and forms based on selfishness, which makes it interpretable. In this spirit we note further that the continuous reinforcement of the *P*s allows each SA-agent to continuously adapt its model of the world and, with that, anticipate/engineer the next state of its environment.

## 2.2 Multiple SA-agents

We study two groups of SA-agents that interact with one another on a 2D plane with periodic boundary conditions. SA-agents of group 1(2) can move with velocities of magnitude $V_1(V_2)$ and can detect the velocity and position of other SA-agents if those are in their vision radius $r_1(r_2)$. If a SA-agent of group 1(2) has some SA-agents of group 2(1) in its vision radius, it receives the payoffs of an anti-coordination game as follows: For a SA-agent in group 1, if $n(>0)$ is the number of SA-agents of group 2 in $r_1$, it receives a payoff of $-n$. Otherwise, it receives 1. For a SA-agent in group 2, if $m(>0)$ is the number of SA-agents of group 1 in $r_2$, it receives a payoff $m$. Otherwise, it receives the payoff -1.

The decision process of the SA-agent used in this study is shown in Figure 2. Each SA-agent can use the information (the velocity and position) of in-group or out-group SA-agents in its vision radius. The decision on which information to use is adaptive for each agent. So, SA-agents have a reliance ($R(t)$) of $P$ values, which is the first level of decision making. If the SA-agent, based on its $R(t)$, decides to use the in-group information, it picks its next direction to be the average of those SA-agents. If there are no in-group SA-agents in its neighborhood, it keeps its previous direction. On the other hand, if the SA-agent decides to use the information of out-group SA-agents, it uses the second level of decision making in which it anticipates the position of those SA-agents and has to decide to pick a direction away from or toward that position based on its oppose-follow ($OF(t)$) $P$ values. Again, if no out-group SA-agents are in its neighborhood, it keeps its previous direction.

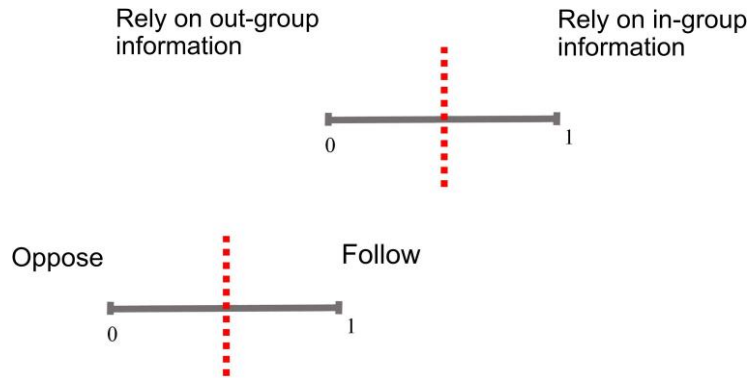

**Fig. 2** Schematics of the two-level binary decision mechanism of SA-Agent used in this work. Each SA-agent makes a decision based on the $P$s of the corresponding choice and the random number generated in the trial. At the first level, the SA-agent decides to either use the information of its neighboring in-group or out-group SA agents (Reliance, $R(t)$ threshold). If its first decision is to use the in-group information it selects its next direction as of the average of those SA-agents. On the other hand, if it decides to use the information of the out-group, it uses the second level of decision making, i.e., it evaluates the next position of those SA-agents and decides either to move away (oppose) or go toward (follow) that point (oppose-follow, $OF(t)$ threshold).

In our simulations, for simplicity, we studied cases where the number of SA-agents of each group is the same $N = N_1 = N_2$, the magnitude of their velocity ($V_1 = V_1 = 0.01$), and their vision radius ($r_1 = r_2 = 0.15$) are equal. The length of the 2D plane, with periodic boundary conditions, is 1. The change in the $P$s, quantified by $\Delta_{i,t}$, is a function of the last two payoffs of agent $i$:

$$\Delta_{i,t} = x \frac{\Pi_i(t) - \Pi_i(t-1)}{|\Pi_i(t)| + |\Pi_i(t-1)|}, \quad (1)$$

where $x = 0.2$ is a positive number that represents the sensitivity of the SA-agent to the feedback it receives from its environment. The quantity $\Pi_i(t)$ is the payoff of the agent $i$ at time $t$. If the value of $P$ becomes less than 0 or more than 1, it is set back to 0 and 1, respectively.

## 3 Results

We analyzed the change of the complexity, measured by MDEA, of the ensemble average of reliance $R(t)$ and $P$s of oppose-follow $OF(t)$ time series and also the order parameter $O(t)$ of the two groups of SA-agents playing the anti-coordination game. The ensemble averages of $R(t)$ and $OF(t)$ thresholds are characteristics of the internal interactions of the system while the $O(t)$ is a measure of the global behavior of the SA-agents. Figure 3 shows two snapshots of the configuration of SA-agents; At the beginning of the simulation (left panel), where the $N = 50$ SA-agents of each group were randomly distributed and directed, and subsequently, after interacting for $t = 10^4$ trials (right panel), swarms of SA-agents of the same groups are formed.

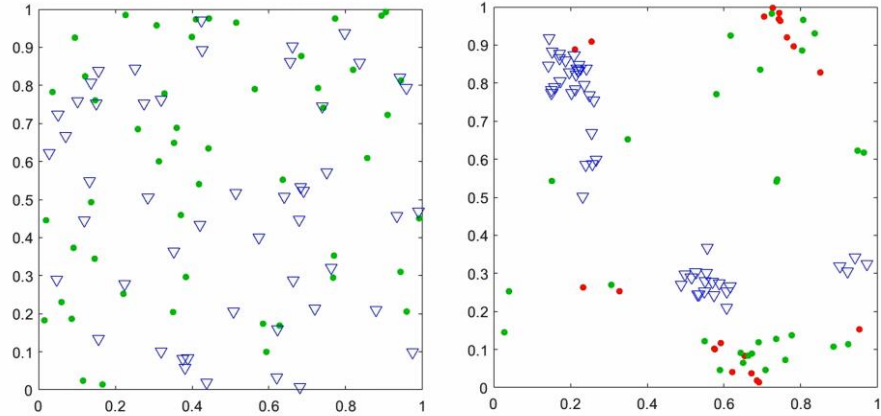

**Fig. 3** Snapshots from the position of the $N = 50$ SA-agents of group 1 (circles) and $N = 50$ SA-agents of group 2 (triangles) playing the anti-coordination game as described in the text. The left snapshot is taken at trial $t = 1$ while the snapshot on the right is taken after $t = 10^4$ trials. The SA-agents of group 1 with $R(t) < 0.25$ are shown with red circles (representing the SA-agents with high propensity ($P >= 0.75$) to rely on in-group information) and otherwise as green circles (for clarity, we haven't distinguished the SA-gents of group 2 in this way.)

Figure 4 shows the evolution of the ensemble averages of $R(t)$ and $OF(t)$ thresholds of the two groups, each with $N = 20$ SA-agents. We stress that these two time series constitute the internal decision making processes of the SA-agents, so we are witnessing the 'cognition' of the SA-agents. Initially $R(1) = 0.5$ and $OF(1) = 0.5$, but over time, the SA-agents learn to adaptively use the information of the in-group and out-group

SA-agents and, if they decide to use out-group information, the SA-agents of group 1 learn to migrate away (cyan curve), while the SA-agents of group 2 learned to migrate towards (orange curve) the anticipated position of the SA-agents of the other group.

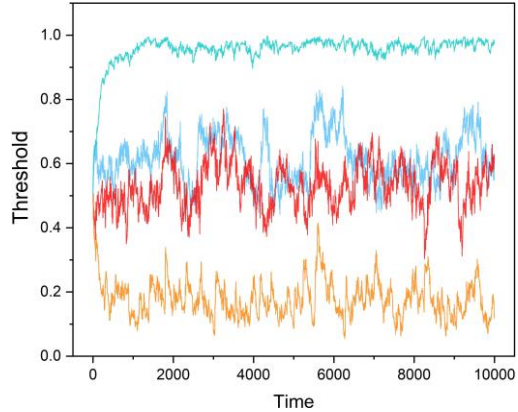

**Fig. 4** The evolution of the ensemble averages $R(t)$ thresholds (blue and red curve respectively for group 1 and 2) and $OF(t)$ thresholds (cyan and orange curve respectively for group 1 and 2). $N = 20$.

Figure 5 shows the scaling indices, resulting from applying the MDEA to the time series of ensemble averages $R(t)$ and $OF(t)$ thresholds of SA-agents of groups 1 and 2. The scaling index time series was measured using slices of $3 * 10^4$ data points (trials) overlapped by $1 * 10^4$ data points. The resulting scalings are found to be time dependent and consequently to be multifractal. It is evident from this figure that scaling parameters for the averaged $R(t)$ thresholds of the two groups are in synchrony, as are the scaling parameters for the averaged $OF(t)$ thresholds. Thus, the scaling behavior of both time series characterizing the SA-agents 'cognition' are multifractal manifesting CS with quasi-periodic scaling. Although not having any mechanism in common with the interacting triad of ONs involving the heart, lungs and brain, the theoretical social SA-agent model depicted in the figure manifest the same CS phenomenon first observed in processing the empirical data from the interacting triad [1, 2].

A third CS is suggested by the vertical bars in this figure introduced to aid the eye in comparing the quasi-periodic variability of scaling time series of averaged $R(t)$ and $OF(t)$ thresholds. Note that the CS of the averaged $R(t)$ thresholds time series appears to be much greater than that of the averaged $OF(t)$ thresholds time series. The potential significance of this difference ought to be explored and we plan to do so in subsequent studies focusing on the parameters of the model.

The cyan diamond symbols in the panels of Figure 6, from top to bottom, show the cross-correlation between the time series of the scaling indices (similar to the ones depicted in Figure 5), evaluated via the MDEA on the averaged $R(t)$, $OF(t)$, and the order parameter $O(t)$ time series of the two groups of SA-agents for different sized groups $N$, respectively. The green squares are the cross-correlations between the corresponding time series (the time series sliced as was done for the MDEA and their cross-correlations averaged). Ten simulations were done for each $N$. While the cross-correlations between the time series remain very low and insensitive to change of $N$, the cross-correlations between the scaling time series are high and measure the intelligence of the system.

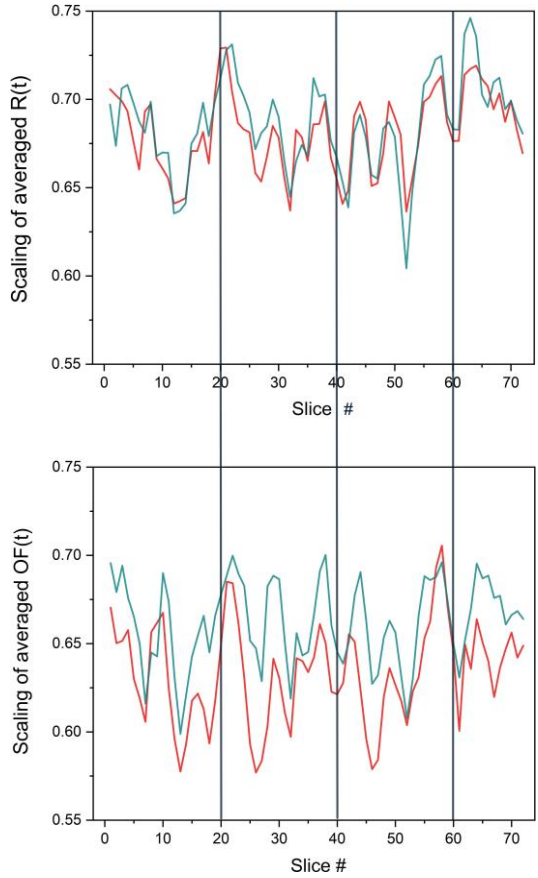

**Fig. 5** The scaling time series of averaged $R(t)$ (top panel) and averaged $OF(t)$ (bottom panel) thresholds time series of the group 1 (cyan) and group 2 (red) SA-agents. $N = 20$. The scalings are evaluated using the MDEA ( stripe size = 0.001) on slices with size $3^4$, moving by $3^4$ data points, where the length of the whole time series was $10^6$. The three vertical lines are to show the synchrony between the top and bottom scaling time series. The cross-correlations of the scaling time series are given in Table 1.

The cross-correlation between the scaling time series of averaged $R(t)$, $OF(t)$ thresholds, and $O(t)$, for $N = 20$, are given in Table 1.

**Table 1** CS between order parameter ($O(t)$, averaged $R(t)$, and averaged $OF(t)$ threshold time series of both groups, $N = 20$, p-values < 0.001.

|       | O-G1   | O-G2   | R-G1   | R-G2   | OF-G1  | OF-G2 |
|-------|--------|--------|--------|--------|--------|-------|
| O-G1  |        |        |        |        |        |       |
| O-G2  | 0.6114 |        |        |        |        |       |
| R-G1  | 0.6247 | 0.5903 |        |        |        |       |
| R-G2  | 0.5560 | 0.5263 | 0.8787 |        |        |       |
| OF-G1 | 0.4729 | 0.3738 | 0.7569 | 0.6781 |        |       |
| OF-G2 | 0.4231 | 0.3811 | 0.6684 | 0.6765 | 0.6944 |       |

Figure 7 shows the averages SA-agent group size $N$. In the top panel of the figure, there is a plateau where the average scaling of the averaged $R(t)$ remains constant for both groups and then starts decreasing at $N = 20$. In the middle panel there is a maximum value for the average scaling of the averaged $OF(t)$ which occurs in different $N$ for each group. In the bottom panel of the figure the average scaling of the $O(t)$ dramatically decreases with increasing group size $N$ for both groups.

The cyan diamonds in Figure 8 show the average (ensemble averages and time averages) payoff of the overall system for each group size $N$. The average payoff increases by increasing $N$ and reaches a maximum of about 0.2 and after that gradually decreases with increasing group size $N$ for both groups. The green squares show the average payoff of the whole system if there were no interaction among the SA-agents. In this case, the average payoff remains very close to zero.

## 4 Discussion

Figure 3 shows that the swarms emerge between SA-agents as they learn from their experience that adaptively changing their $R(t)$ and $OF(t)$ thresholds is beneficial. Note that without interaction between the SA-agents, the payoff of the overall system is close to zero for all values of $N$. On the other hand, for the interacting SA-agents, the sum of the payoff is a positive value (Figure 8.) The time series of the averaged $R(t)$ and $OF(t)$ thresholds of the SA-agents represents the interactions in the network while the $O(t)$ shows the organization's output. We analyzed the complexity of these time series for both groups of SA-agents using MDEA.

In the top panel of Figure 6 the cross-correlation between the scaling time series of averaged $R(t)$ of the two groups is > 0.95 for $N = 10$ and is diminished with increasing group size $N$. In the middle panel of this figure the cross-correlation between the scaling time series of averaged $OF(t)$ of the two group increases in value non-monotonically until reaching a maximum of about 0.6 at $N = 30$ and then non-monotonically decreases with increasing group size $N$. In the bottom panel of this figure the cross-correlation between the scaling time series of the order parameter $O(t)$ of the two groups is similar to that of the averaged $R(t)$ but lower in intensity and diminishes at smaller $N$. Unlike the first two panels, which depict the influences on the decision-making process, the order parameter depicts the results of the decision-making on the average behavior of the SA-agent.

The results show that these time series have anomalous scaling in the range of (0.5, 1). More importantly, our results show that the scaling of these time series varies in time in a way that are in synchrony with one another (see Figure 5, Figure 6, and Table 1). This synchrony phenomenon was recently found in the time series of EEG, ECG, and Respiratory data [1, 2], wherein the name Complexity Synchronization (CS) was coined. As Figure 6 shows, CS can distinguish between different systems of size $N$ while the typical cross-correlation between the time series is very low and is not sensitive to changes of $N$.

The panels of Figure 7 show that the average of the scaling time series of these time series also depends on $N$. These results suggest that CS is a general property of mutual adaptive environments that can be used to quantify EI at both the levels of the average of the time series scaling index and the magnitude of their cross-correlations.

The social model we used for CS in EI is a MABM utilizing SA-agents. Rather than discussing the details of the EI using this model for different parameters, we used it to show that CS is a natural outcome of self-organization. This social model encourages future theoretical work on CS in a totally different context from that in which it was discovered. Note that our social MABM should not be confused with models where the dynamics of the system rely on a macroscopic control parameter that must be set by the experimenter. For example, in the Vicsek model [22], the uncertainty in measuring the average velocity by the agent is introduced in the model as noise, which plays the role of a control parameter, and in [40], the selection strength $β$ is the macroscopic control parameter. In our social MABM, there is no control parameter; instead, the other SA-agents constitute the environment of the SA-agent of interest and they replace the role of control with a bottom-up self-organizing process originating from the self-interest of the SA-agents.

## 4.1 Emergence of CEs

Viswanathan et. al in [41] studied the foraging behavior of a prototypical wandering albatross and found an IPL distribution of flight time intervals with $μ = 2$, indicating its flight path is a Lévy process. They interpreted the existence of such temporal scale invariance to be a consequence of a possible scale-invariant spatial distribution of food resources for the albatross. This suggested that we focus our work on the behavior of a single SA-agent and because of its interactions with other SA-agents, its dynamics constitute a Lévy Walk.

The top panel of Figure 9 shows the trajectory of a single SA-agent of group 1 where it is interacting with the other 19 in-group and 20 out-group SA-agents ($N = 20$). The advantage of studying a single SA-agent's trajectory is that the events are clearly marked as the discrete times the SA-agent changes its direction in the one-dimensional projection of that trajectory. The projection of the 2D trajectory of the SA-agent onto the X-axis assists us in illustrating how in processing empirical datasets stripes extract events.

The middle panel of Figure 9 shows the velocity of the same SA-agent projected onto the X-direction $V_x(t)$, which is indeed the projection of the 2D trajectory shown in the top panel. The three red-dotted lines in the middle panel divide it into four stripe regions, and every time the $V_x(t)$ passes from one stripe region to another we interpret that crossing as an event (marked as an event of positive unit amplitude), as depicted in the bottom panel. This creates a discrete/binary representation of the dynamics of the original time series. In the MDEA, we use these randomly spaced events to create the steps for a diffusion process. Note that the bottom panel of this figure is equivalent to panel b of Figure 2 in [1]. In the case of real data, it is often the average of many signals, which, using stripes, we can recover events and use them to carry out the entropy analysis of the empirical diffusion process.

Figure 10 shows the average scaling of single trajectories for different $N$. The values of the scaling parameter all fall in the interval (0.65, 0.9), which indicates anomalous diffusion statistics. Also, in the supplementary material, we show using the aging experiment [42] that the events are renewal, indicating the emergence of crucial events with statistics similar to those of a Lévy Walk. Note, however, that the scaling in our model changes over time rather than being fixed, as it is in the case of a simple Lévy Walk. This change from a constant to a time-dependent scaling parameter depicts the transition from a monofractal to a multifractal process. These results show that in the social MABM a Lévy PDF is an emergent property of mutual interaction between the two competing groups of SA-agents.

Here we note preliminary calculations indicate that using detrended fluctuation analysis (DFA) fails to reveal the existence of CEs in the data. If this conclusion is borne out by future analysis, it Would have profound implications, not the least of which would be the failure to identify CS and what that entails.

## 4.2 Implications of CS

Implications of this research are dramatic and far-reaching in that the theoretical and analytical approaches may generalize across a broad spectrum of questions from complex human neurophysiological ONs to complex social networks of humans, as well as into the realm of technology networks. We are pursuing a more comprehensive research plan to translate complexity science concepts, methods, and tools to seek organizing and operating principles valid across both biological and sociological networks.

Given that the CS among the time series generated by SA-agents in a social network is comparable to the CS among time series generated by ONs within a NoONs, a successful extension of cognition via human-EI interaction is implied. An advantage of human-EI interaction is that EI, as an adaptive decision-making machine, logs/refreshes its interpretation of its dynamic environment in $P$ sets, forming multi-layer adaptive networks. So, these networks are naturally created using fuzzy logic and are rooted in self-interest. Consequently, in contrast with the black box of AI networks, the networks of EI are interpretable. This makes EI an appropriate candidate to model biological networks that operate based on electrochemical energy (e.g., firing thresholds of neurons), which may be analogous to reinforcement learning in social-environmental networks.

## 5 Conclusion

Little is known about how to observe, manage, and improve biological/artificial/hybrid human-machine networks. In previous work, we showed that by examining biological datasets at the level of their scaling indices, there is synchrony among their complexity indices. In particular, the investigation presented herein establishes that although the MABM social model has no mechanism in common with that of the interacting triad of ONs involving the heart, lungs and brain, the theoretical social SA-agent model manifests the same CS phenomenon first observed in processing the empirical data from the interacting triad of ONs [1, 2].

Thus, we conjecture that just as all linear dynamic systems that are periodic in time, whether their periodicity is a consequence of a spinning wheel, an equidistant set of points along a line, the sound of single frequency musical note, or any of a large number of other physical processes, can be described using a simple harmonic oscillator, so too can all complex dynamical systems having CS in time, whether the CS is a consequence of multiple interacting ONs or interacting social groups of arbitrary size, can be described by their synchronously locked multifractal dimensions. We conjecture further that the fractal dimension in a complex network plays a role analogous to that of the frequency in a simple linear system. Moreover, a multifractal dimension in a complex network plays a role analogous to that of a frequency spectrum in complicated but dynamically linear system.

The present work uses a relatively simple MABM social model incorporating SA-agents into the network dynamics to show that CS is a general property of mutually adaptive environments and that the emerging CS between SA-agents' time series is compatible with the CS in human biological/behavioral data. The ensemble average multifractal dimension $<D(t)>$ is herein given by $<D(t)> = 2 - R(t)$ and therefore we found in the Results section that $R(t)$ has a quasi-harmonic variability in the relatively narrow interval [1.26,1.4]. This synchronous behavior displayed by $R(t)$ in Figure 5 is not that different from the quasi-periodic scaling of the multifractality of $OF(t)$ parameter depicted there as well.

These findings suggest that the EI of SA-agents is a candidate for modeling the interactions with humans for training/teaming/rehabilitation. Also, it is worth emphasizing the difference between the present modeling strategy and that involving artificial intelligence (AI) by pointing out that the EI of SA-agents is based on fuzzy logic and self-interest in contrast to AI being based on learning from selected human datasets and is without self-interest. This makes EI interpretable for humans, while the AI is considered a black box. Note that there is no macroscopic control parameter in the EI of SA-agents with control being a bottom-up emergent property of the underlying dynamic process.

**Code availability.** All the codes used to produce the results of this work are available at https://github.com/Korosh137/Complexity-Synchronization-in-Emergent-Intelligence.

**Data availability.** The data used to produce the figures are available at https://drive.google.com/drive/folders/1uZPi e1T4 3DHh8ZAT1bct300m1U2z0B? usp=sharing.

**Acknowledgments.** Research was sponsored by the Army Research Laboratory and was accomplished under Cooperative Agreement Number W911NF-23-2-0162. The views and conclusions contained in this document are those of the authors and should not be interpreted as representing the official policies; either expressed or implied, of the Army Research Laboratory or the U.S. Government. The U.S. Government is authorized to reproduce and distribute reprints for Government purposes notwithstanding any copyright notation herein.

**Author contributions.** K.M conceived CS in EI and conducted the analysis. K.M and B.J.W wrote the draft of the manuscript. All authors critically assessed and discussed the results, and revised and approved the manuscript.

**Competing interests.** The authors declare no competing interests.

# Appendix A    Renewal experiment for the events of an SA-agent's trajectory

To check whether or not the events extracted using stripes are renewal (i.e. consecutive events are statistically independent of one another), we use the renewal experiment (RE) introduced in [42]. The first step in the RE is aging the events with a given time $t_a$. For aging, as shown in Figure A1, we start from the events, and the time distance to the next event is the aged event. Collecting the aged time intervals, we can evaluate their waiting-time PDF. Note that aging influences small events more than large ones, and after normalization, the weight of short time intervals decreases while the weight of the large time intervals increases. Also, note that aging removes some of the short time events.

To check that the extracted events are renewal, i.e., independent consecutive events, we first shuffle the original $\tau_i$ and then age them and finally evaluate the waiting-time PDF of the resulting time intervals. If the waiting-time PDF of the two cases of aged and shuffled-aged are the same, it shows that there is no correlation between the events. Figure A1 shows the waiting-time PDF of the events, extracted using stripes, for the $V_x(t)$ before aging, aged, and shuffled-aged. There is IPL in the waiting-time PDF of all tree cases. Also, the waiting-time PDF of the aged and shuffled-aged events are the same, which shows the events are renewal, confirming that the dynamics of the single SA-agent is ruled by the renewal process (crucial events). Note that the evaluated IPL of the waiting-time PDF of the extracted events $\mu$ in Figure A1 is related to $\delta$ in Figure 10 by equation $\mu = 1 + 1/\delta$.

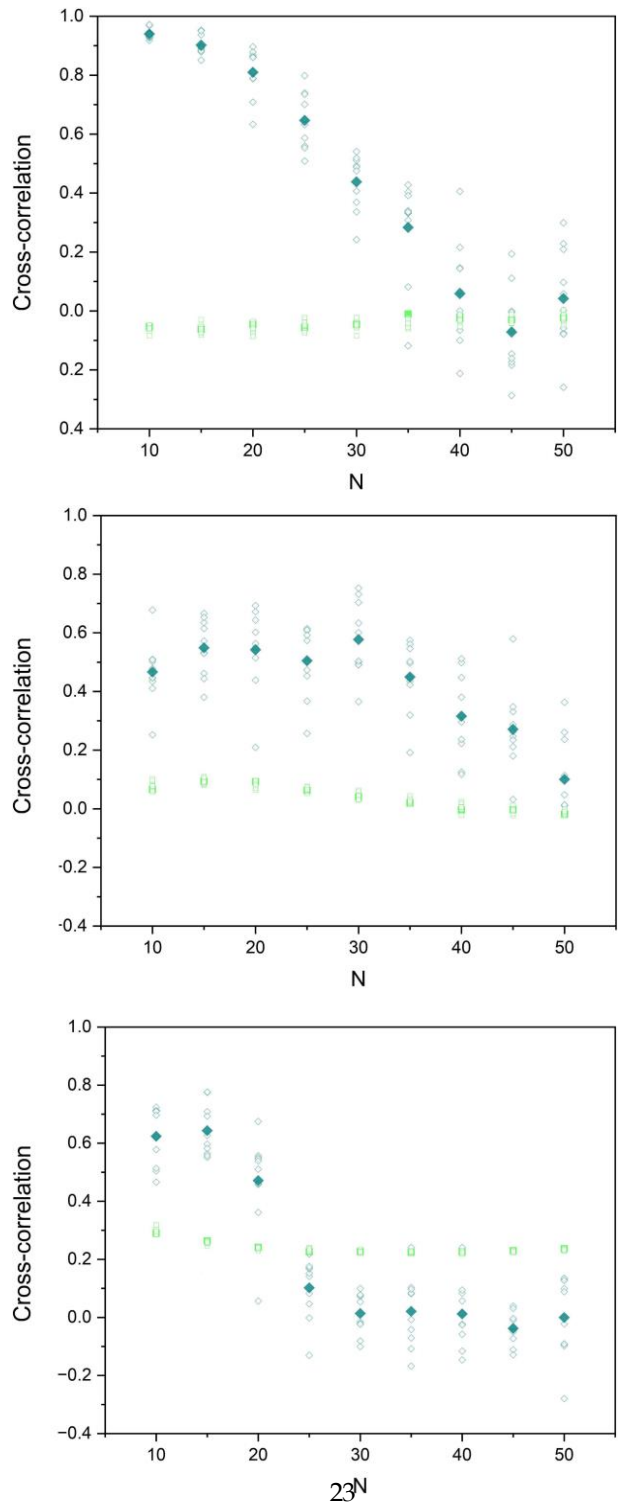

**Fig. 6** In the panels, from top to bottom, the cyan diamonds show the cross-correlation between the scaling time series evaluated using MDEA (with stripe size = 0.001) on averaged $R(t)$, $OF(t)$ thresholds, and $O(t)$ time series of the two groups, vs. the number of SA-agents of each group $N$. The green squares show the average cross-correlation between the corresponding slices of data that were used for MDEA. The filled symbols show the mean values of the cross-correlations of the ten simulations done for each size $N$.

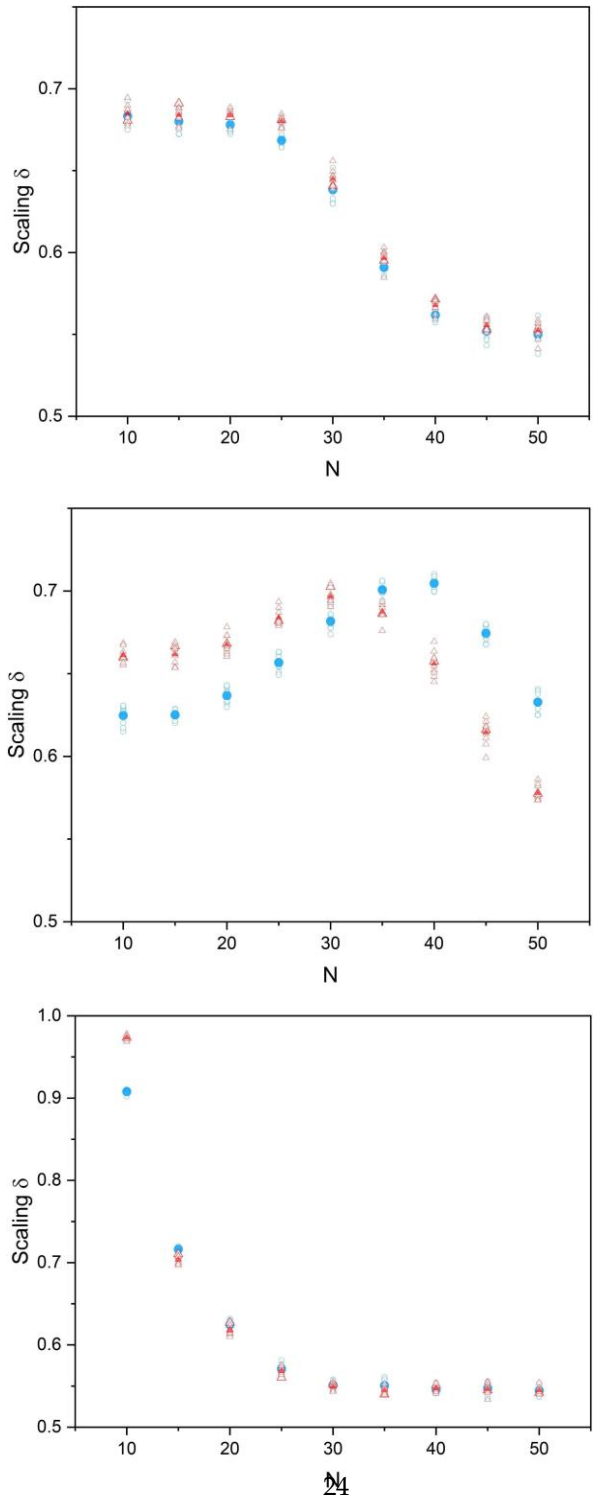

**Fig. 7** The panels, from top to bottom, show the average of the scaling time series, evaluated using MDEA (with stripe size = 0.001) on the averaged $R(t)$, $OF(t)$ thresholds, and $O(t)$, vs. the number of SA-agents of each group $N$. The blue circles and red triangles represent groups 1 and 2, respectively. The filled symbols show the mean values of the ten simulations done for each size.

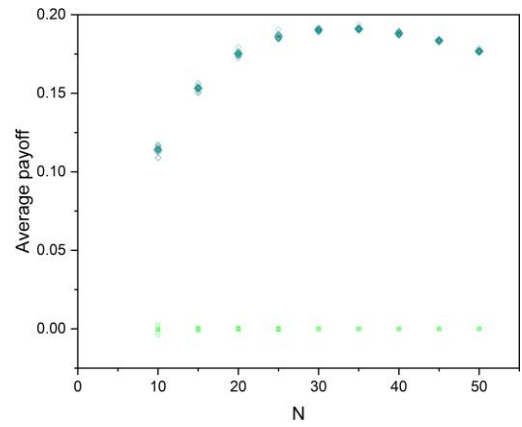

**Fig. 8** The cyan diamonds show the time average of the total payoff of the networks for different group size $N$. The green squares show the time average of the total payoff of the same systems in the absence of interaction between the SA-agents. The filled symbols show the mean values of the ten simulations done for each size.

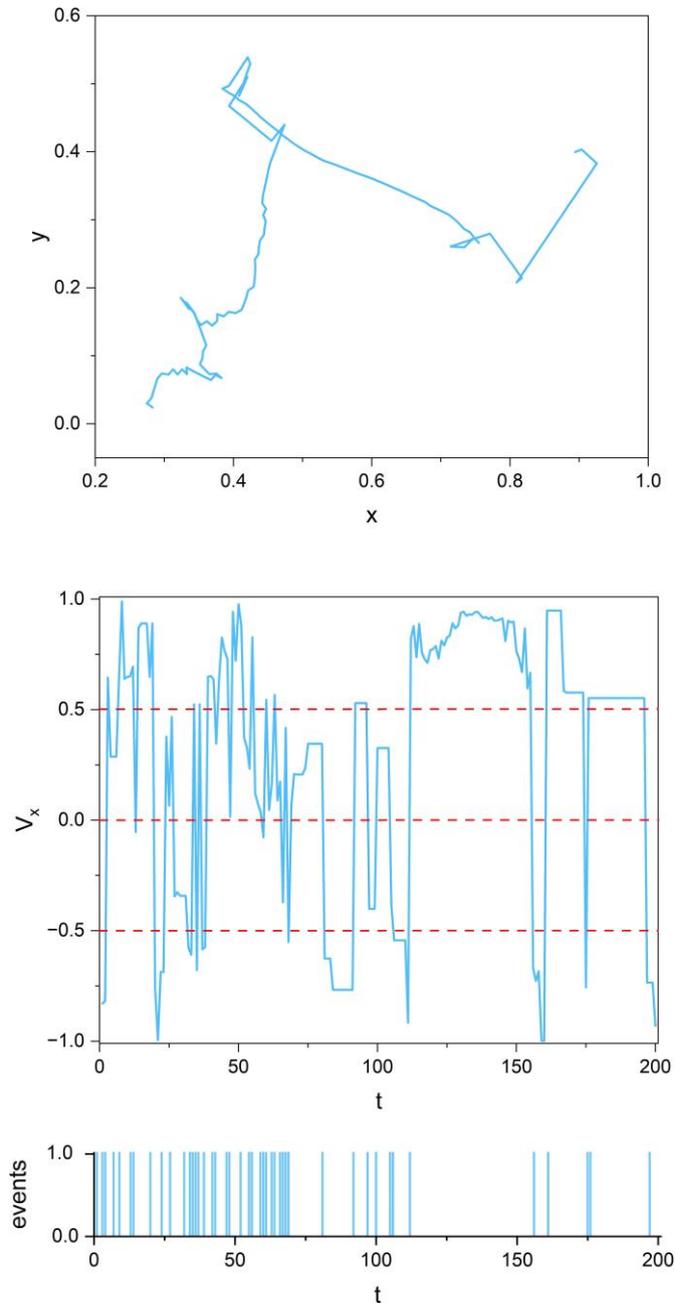

**Fig. 9** Schematics of the use of stripes in extracting events as a discrete/binary representation of the data. Top panel: part of the 2D trajectory of one SA-agent of group 1 (out of 19 group 1 and 20 group 2 interacting SA-agents). Middle panel: the time series of the velocity of the agent (top panel) projected onto the X-axis, $V_x$. The horizontal red dotted lines depict four stripes, and every time the $V_x$ time series passes across one stripe region into another, it is marked as an event (the blue bars in the bottom panel.

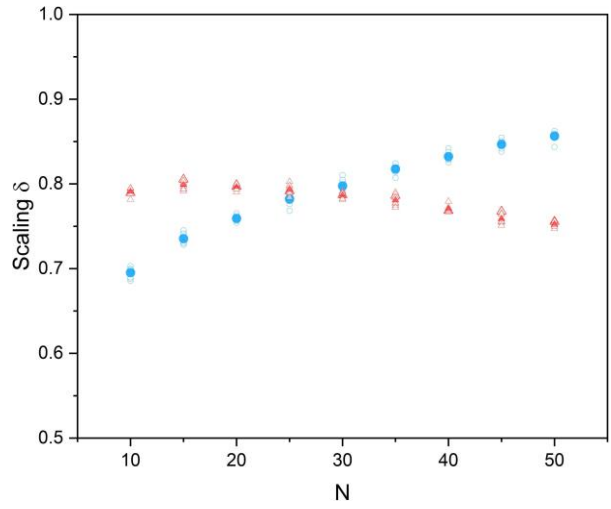

**Fig. 10** The average scaling of a single SA-agent's trajectory for different $N$. The blue circles and red triangles represent groups 1 and 2, respectively. The filled symbols show the mean values of the ten simulations done for each $N$.

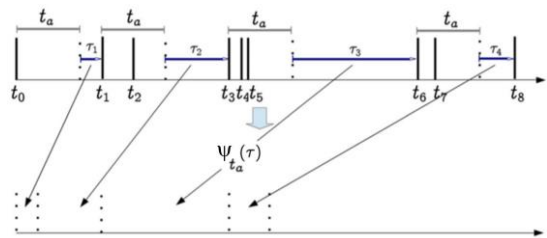

**Fig. A1** Schematics for the aging experiment. The time interval $\tau_i$, extracted using stripes aged by $t_a$. From [15] with permission.

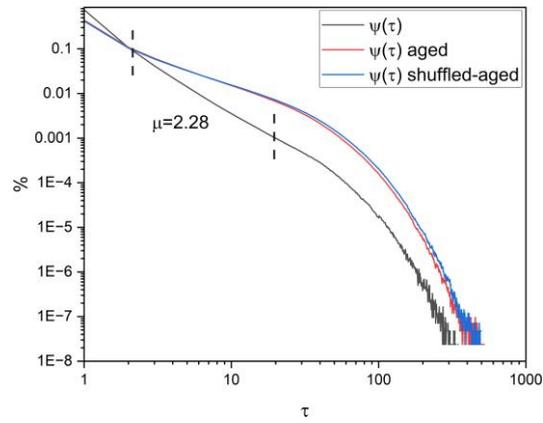

**Fig. A2** Renewal test for the events extracted using stripes from the velocity of the projection of the 2D trajectory of a single SA-agent from group 1, onto the x-axis $V_x(t)$. The black curve shows the waiting-time PDF of the extracted events, with IPL of $\mu = 2.28$. The red and blue curves are the waiting-time PDF of the same events after being aged and being shuffled-aged, respectively. $N = 20$. Stripe size = 0.01, $t_a$=100, length of the time series = $10^7$ trials.